# Robust Ferromagnetism in Silicene Nanoflakes through Patterned Hydrogenation

S. M. Aghaei [a], I. Torres [a], and I. Calizo [*a, b]

[a] Quantum Electronic Structures Technology Lab, Department of Electrical and Computer Engineering, Florida International University, Miami, Florida 33174, United States
[b] Department of Mechanical and Materials Engineering, Florida International University, Miami, Florida 33174, United States
[*] E-mail: icalizo@fiu.edu

**Abstract**

Considerably different properties emerge in nanomaterials as a result of quantum confinement and edge effects. In this study, the electronic and magnetic properties of quasi-zero-dimensional silicene nanoflakes (SiNFs) are investigated using first-principles calculations. While the zigzag-edged hexagonal SiNFs exhibit non-magnetic semiconducting character, the zigzag-edged triangular SiNFs are magnetic semiconductors. One effective method of harnessing the properties of silicene is hydrogenation owing to its reversibility and controllability. From bare SiNFs to half-hydrogenated and then to fully hydrogenated, a triangular SiNF experiences a change from ferrimagnetic to very strong ferromagnetic, and then to non-magnetic. Nonetheless, a hexagonal SiNF undergoes a transfer from non-magnetic to very strong ferromagnetic, then to non-magnetic. The half-hydrogenated SiNFs produce a large spin moment that is directly proportional to the square of the flake's size. It has been revealed that the strongly induced spin magnetizations align parallel and demonstrates a collective character by large range ferromagnetic exchange coupling, giving rise to its potential use in spintronic circuit devices. Spin switch models are offered as an example of one of the potential applications of SiNFs in tuning the transport properties by controlling the hydrogen coverage.

## 1. Introduction

The discovery of graphene in 2004, along with its exotic properties provided new insight in the field of condensed matter science [1]. Toxicity, large area growth, and incompatibility with current silicon-based nanoelectronics are the main challenges for graphene-based device production, inspiring a search for other two-dimensional (2D) nanostructures to produce fast nanoscale electronic devices that do not require retooling. Silicene as a 2D material composed of a group-IV element silicon) has fascinated the community not only for being easily integrated into Si-based technology [2], but also because of its superlative properties including ferromagnetism [3], half-metallicity [4], quantum Hall effect [5], giant magnetoresistance [6], and superconductivity [7]. In 1994, Takeda and Shiraishi [8] investigated silicene in a theoretical study for the first time, albeit the name silicene was coined by Guzmán-Verri and Lew Yan Voon in 2007 [9]. Experimental synthesis of monolayer silicene with conclusive results was realized by Vogt *et al.* in 2014 [10]. Silicene has been experimentally synthesized on different substrates such as Ag [7, 10-12], Ir [13], $ZrB_2$ [14], ZrC [15], and $MoS_2$ [16], for various phases. Several potential applications of this novel material have been proposed in the area of spintronics [17], field-effect transistors (FETs) [18], and sensing devices [19, 20]. Investigation of electronic structure of silicene through angle-resolved photoemission spectroscopy (ARPES) and scanning tunneling spectroscopy (STS) measurements proved that it is a zero band gap semimetal [10, 21]. In order to induce a band gap in a silicene sheet, various procedures have been studied, such as doping [22], substrate effects [23], chemical functionalization [24, 25], electric field [17], silicene nanoribbons (SiNRs) [26, 27], and introducing nanoholes [28, 29].

The tendency of silicon atoms to adopt $sp^3$ and $sp^2$ hybridization over only $sp^2$ hybridization [10] makes its honeycomb structure buckled [9, 30], distinguishing it from graphene which is a flat sheet. Hence, silicene is naturally more favorable for atom and molecule adsorption [31-34] than graphene, resulting in a great deal of applications in the area of hydrogen storage [31], thin film solar cell absorbers [35], hydrogen separation membranes [32], and molecular sensors [33, 34]. Hydrogenation was found as a favorable chemical method to modify the electronic and magnetic properties of graphene because of its reversibility [36] and controllability [37, 38]. There have been quite a number of theoretical [3, 4, 39-56] and experimental [57-59] studies on the hydrogenation of silicene in the literature. Intriguing properties in hydrogenated silicene have been reported through theoretical calculations, for example, large gap opening [39], and extraordinary ferromagnetic (FM) [4] and optoelectronic properties [44]. A fully hydrogenated derivative of silicene, known as silicane which is the silicon counterpart of graphane, has been expected to have a band gap in the range of 2.9–3.8 eV, depending on the configuration [41]. Ergo, it would be considered a potential material for optoelectronics and field effect transistor (FET) applications [46]. The hydrogenation process in silicene is uniformly ordered and reversible, making it easier to manipulate the hydrogen coverage [57]. Consequently, the half-hydrogenation can be achieved by hydrogenating of silicene from one side while keeping the other side intact similar to its carbon counterpart graphone. It was found that half-hydrogenated silicene shows FM semiconducting behavior with a band gap of 0.95 eV [4, 56]. Besides, a long-range room temperature FM coupling between Si atoms can be realized in half-hydrogenated silicene [56]. Osborn *et al.* have theoretically shown that H atoms prefer to be adsorbed in pairs to form the most stable configuration of partially hydrogenated silicene [40]. By altering the concentration of hydrogen atoms on silicene, the characteristic of its band structures can be varied from metallic to magnetic semiconducting and then to non-magnetic (NM) semiconducting [43].

As nanomaterials approach lower dimensions, their properties noticeably change due to the quantum confinement and edge effects. The quasi-one-dimensional (1D) nanoribbons show semiconducting or metallic behavior depending on their edge types and the width of the ribbon [25, 60]. Further confinement of silicene has sparked interest in the study of quasi-zero-dimensional (0D) silicene nanoflakes (SiNFs) since understanding their properties is of great significance in order to continue advancing nanoscale electronic and spintronic device fabrication. Luan *et al.* found that electronic and magnetic properties of SiNFs depend strongly on their size and shape [61]. The hexagonal zigzag SiNF exhibits NM semiconducting behavior, while the triangular zigzag SiNF is a magnetic semiconductor.

Although hydrogenated silicene sheets have been recently discovered, the effects of hydrogen adsorption on SiNFs have not been explored yet. In this paper, we employed first-principles methods based on density functional theory (DFT) to investigate the electronic and magnetic properties of SiNFs with different configurations. The patterned adsorption of hydrogen on SiNFs is systematically explored, and its stability is evaluated by calculating the formation energies. Contrary to bare SiNFs, the hydrogenated SiNFs show unique magnetic properties. This study also proposes a simple and efficient approach to engineering the transport properties of electrons by controlling the coverage of hydrogenation on SiNRs, which may pave a new path to explore spintronics at the nanoscale.

## 2. Computational Methods

Our first-principles calculations are performed based on DFT combined with nonequilibrium Green's function (NEGF) implemented by Atomistix ToolKit (ATK) package [62-64]. The hybrid B3LYP exchange-correlation functional is utilized as the exchange and correlation functional to solve Kohn-Sham equations. The electronic wave functions are expanded by a double-$\zeta$ polarized basis for different atoms, while the cut-off energies are chosen to be 600 eV. To eliminate the interactions between adjacent nanoflakes, a 30 Å vacuum slab is considered in all directions. The electronic temperature is kept constant at 300 °K during calculations. For geometry optimization, both supercell and the atomic positions are allowed to be fully relaxed until the force and stress are less than 0.0025 eV/Å and 0.005 eV/Å$^3$, respectively.

## 3. Results and Discussions

### 3.1. The structure, electronic, and magnetic properties of SiNFs

In this study, four types of hydrogen passivated SiNFs with different shapes (hexagonal and triangular), edges (armchair and zigzag), and sizes are studied. In order to differentiate the various configurations of SiNFs, the abbreviations are used as follows: *AH* stands for armchair-edged hexagonal, *ZH* represents zigzag-edged hexagonal, *AT* implies armchair-edged triangular and *ZT* signifies zigzag-edged triangular. The AT-SiNF and AH-SiNF can be further categorized by the integer number dimer lines across one side of the nanoflake, while the ZT-SiNF and ZH-SiNF are identified by the integer number of edge Si atoms across one side of the nanoflake. The model structures of 3-AH-SiNF, 4-AT-SiNF, 3-ZH-SiNF, and 4-ZT-SiNF, as representatives of the aforementioned configurations, are presented in Fig. 1, in which all structures are made of freestanding silicene and all Si edge atoms are terminated by H atoms. The cyan and red balls signify Si and H atoms, respectively. After structural relaxation, the buckling height between sublattices A and B is ~ 0.50 (0.51) Å in Z-SiNFs (H-SiNFs). The distance between Si-H atoms is found to be 1.50 Å, indicating that H atoms are chemically attached to SiNFs' edge. The edge Si-Si bond lengths are 2.23-2.24 Å for H-SiNFs and 2.23-2.25 Å for Z-SiNFs, which are less than the bond length of inner Si-Si (2.28 Å) (similar to pristine silicene [25, 65]), indicating relatively large atomic distortion at the edges.

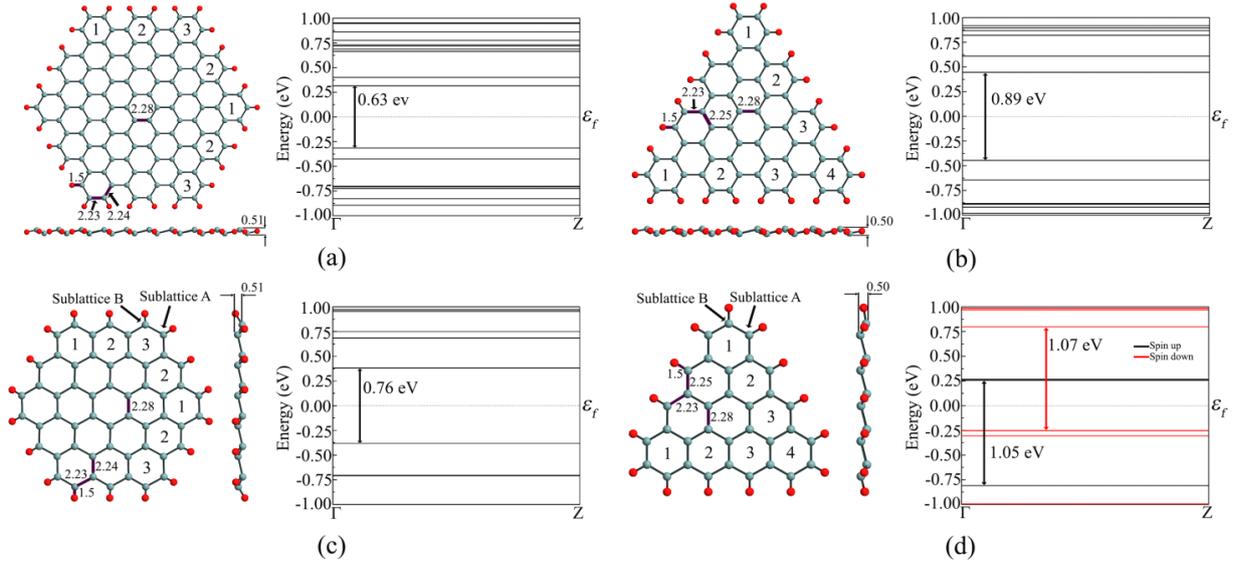

Fig. 1 Schematic structural models of (a) 3-AH-SiNF, (b) 4-AT-SiNF, (c) 3-ZH-SiNF, and (d) 4-ZT-SiNF as representatives of armchair-edged hexagonal SiNF, armchair-edged triangular SiNF, zigzag-edged hexagonal SiNF, and zigzag-edged triangular SiNF, respectively, and their corresponding band structures. The cyan and red balls represent Si and H atoms, respectively. The bond lengths and the buckling distances (in unit of Å) are also given. Black and red lines are spin-up and spin-down channels in (d), respectively.

The band structures of 3-AH-SiNF, 4-AT-SiNF, and 3-ZH-SiNF are also shown in Fig. 1. Our calculations show that N-AH-SiNFs, N-AT-SiNFs, and N-ZH-SiNFs are NM and exhibit semiconducting behavior. Unlike the band structure of a silicene sheet in which the bonding $\pi$ and antibonding $\pi^*$ band cross at $K$ points in the Brillouin zone [9, 29], finite gaps of 0.63, 0.89, 0.76 eV are opened between the valence band maximum (VBM) and the conduction band minimum (CBM) of 3-AH-SiNF, 4-AT-SiNF, and 3-ZH-SiNF, respectively. The energy band gap in SiNF is attributed to the quantum confinement and edge effects caused by the reduced dimensionality of the structures. As displayed in Fig. 2, the band gaps decrease gradually by increasing the nanoflakes size. These results are in agreement with previous theoretical findings [61].

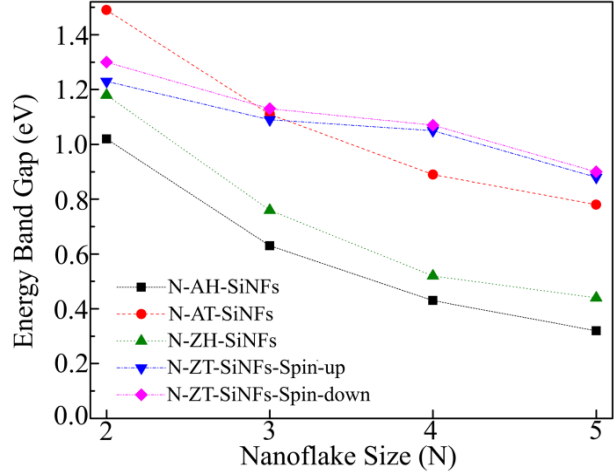

Fig. 2 N-AH-SiNFs, N-AT-SiNFs, N-ZH-SiNFs, and N-ZT-SiNFs energy band gap dependence on nanoflakes size. The energy band gaps decrease as the size of the nanoflakes increase.

Our spin-polarized calculations demonstrate that the ground state of N-ZT-SiNFs have magnetic behaviors. Fig. 1 shows the band structure of 4-ZT-SiNF as a representative of N-ZT-SiNFs. As can be seen, the $E_f$ lies close to VBM of the spin-up channel, and CBM of the spin-down channel. The value of band gap for the spin-up channel (1.09 eV) is less than that of the spin-down channel (1.30 eV). The isosurfaces of spin density distribution ($\rho_{spin-up} - \rho_{spin-down}$) of 4-ZT-SiNF is plotted in Fig. 3(a). The system provides a ferrimagnetic spin ordering with opposite spins on Si atoms of sublattices A ($Si_A$) and B ($Si_B$) in which the total spin-up is larger than the total spin-down. The energy difference between spin-polarized and spin-unpolarized states of 4-ZT-SiNF is 0.29 eV, in favor of spin-polarized state. The spin density of the inner Si atoms is less than the edge Si atoms. Besides, the Si atoms at three corners of nanoflakes, located in sublattice B, have opposite spin direction compared to the outermost Si atoms. This may originate from the charge redistribution of Si-3p states in the region near the edges where the Si-Si bond lengths shrink by 0.05 Å. The total magnetic moment found on our calculations for 3-ZT-SiNF is 2.0 $\mu_B$, and for 4-ZT-SiNF is 3.0 $\mu_B$, showing that the total magnetic moment (M) of N-ZT-SiNF is equal to $N-1$. It indicates that our results are in excellent agreement with Lieb's theorem for bipartite lattices [66]. To visualize the electron redistribution of ZT-SiNFs, the electron difference density of 4-ZT-SiNF is presented in Fig. 3(b). One could see that electrons accumulated on H atoms and between neighboring Si atoms. The high electron density between Si-Si shows this fact that electrons are shared by Si atoms, resulting in the formation of a covalent bond between them. Furthermore, since Si is less electronegative than H, the Si-H bond is polarized toward H, resulting in negative charging of H atoms. The results are also similar to the corresponding cases for graphene [67, 68].

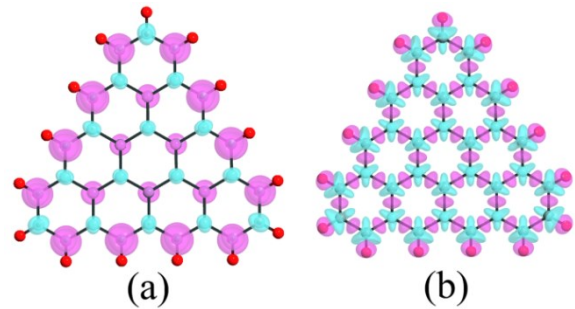

Fig. 3 (a) Spin-charge density distribution ($\rho_{spin-up}-\rho_{spin-down}$) and (b) electron difference densities of 4-ZT-SiNF. The cyan and red balls represent Si and H atoms, respectively. The pink and blue regions correspond to the isosurfaces of spin-up and spin-down states. The isosurfaces value is 0.012 a.u.

### 3.2. Adsorption of a single hydrogen atom on SiNFs

To understand the effect of hydrogenation on the electronic and magnetic properties of SiNFs, the adsorption of a single H (represented by *sH*) atom on 4-ZT-SiNF and 3-ZH-SiNF (as representatives of zigzag-edged SiNFs) are analyzed. It is found that $Si_B$ is energetically more favorable than that of $Si_A$ for hydrogen adsorption. The difference of adsorption energies between two sublattices for 4-ZT-SiNF is 0.9 eV, and for 3-ZH-SiNF is 0.2 eV, in support of $Si_B$. Fig. 4(a) shows the geometric structures of sH-4-ZT-SiNF and sH-3-ZH-SiNF in which a single H atom is adsorbed on the $Si_B$ of the nanoflakes. The Si-H bond length is 1.5 Å, indicating a strong Si-H covalent bond. A charge transfer from Si to H atom occurs, as shown in Fig. 4(b), where electrons are piled up on H (pink area) and the electron density on hydrogenated $Si_B$ (blue area) is slightly increased. Also, the electron orbital overlap can be observed in the $Si_B$-H. Such a charge transfer makes the nearest $Si_A$-$Si_B$ bonds to the H atom longer (weaker) but the second nearest $Si_A$-$Si_B$ bonds shorter (stronger), where the lengths are 2.35 Å and 2.25 Å, respectively. Furthermore, the buckling height between two sublattices where H atom is adsorbed is 0.71 Å which is much greater than that of bare SiNF (0.50 Å). The bond angle of $Si_A$-$Si_B$-H is about 110° which is close to tetrahedral molecule geometry. These values are consistent with earlier research on silicene [52, 53].

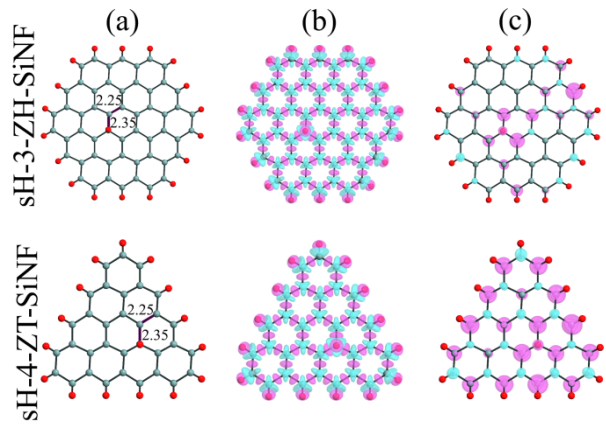

Fig. 4 (a) Optimized geometric structures (b) the corresponding electron difference densities and (c) the corresponding spin-charge density distribution of sH-3-ZH-SiNF and sH-4-ZT-SiNF.

As explained above, the 4-ZT-SiNF is magnetic, and 3-ZH-SiNF is NM with spin moments of 3.0 and 0.0 $\mu_B$, respectively. It is found that the magnetism is induced in 3-ZH-SiNF which is inherently NM and enhanced in 4-ZT-SiNF upon adsorption of single H atom. Fig. 4(c) shows the isosurfaces of the spin density distribution of sH-4-ZT-SiNF and sH-3-ZH-SiNF. As can be seen, with adsorption of a single H on $Si_B$, the spin-down charges on $Si_B$ are faded; interestingly certain spin-up charges appear on H atom, and the spin-up charges on three $Si_A$ neighboring the hydrogenated $Si_B$ are slightly increased. As a result, the total spin moments are augmented, for sH-4-ZT-SiNF increases to 4.0 $\mu_B$, and for sH-3-ZH-SiNF rises to 1.0 $\mu_B$. It means that a single H atom adsorbed on $Si_B$ can increase the total spin moment of the SiNFs by 1.0 $\mu_B$. This phenomenon can be explained as follows. In addition to the strong σ bond between $Si_A$-$Si_B$ atoms in the bare SiNF, a weak π bond exists between unpaired and localized $p_z$ electrons of the $Si_A$-$Si_B$ atoms. When a single H atom approaches $Si_B$, the quasi π bond is destroyed, and true $sp^3$ hybridization occurs by forming a strong σ bond between $Si_B$-H and a weaker $Si_A$-$Si_B$ bond. Furthermore, the $p_z$ electrons of $Si_A$ are left unpaired and delocalized, causing them to move around the $Si_A$ and the H atom. As a result, the spin-up charges accumulate on H atom and $Si_A$, and the $Si_B$ is depleted from spin-down charges. The contribution of $Si_A$ to magnetism is much larger than that of H atom, as will be explained later. It is worth mentioning that when the H atom is conversely adsorbed on Si$_A$, the total spin moment of sH-4-ZT-SiNF and sH-3-ZH-SiNF become 2.0 and −1.0 μ$_B$, respectively, showing a decrease in the total spin moment by 1.0 μ$_B$.

### 3.3. Higher hydrogen coverage on SiNFs

As previously mentioned, each H atom induces a spin moment of 1.0 μ$_B$ in SiNF. Therefore, higher H concentration is desired to further enhance the magnetism in the SiNFs. To this end, half-hydrogenated (hH) SiNFs, in which all Si$_B$ atoms are hydrogenated, and all Si$_A$ atoms remain unsaturated are considered. Fig. 5(a) shows the optimized geometries of hH-4-ZT-SiNF and hH-3-ZH-SiNF. The relaxed bond lengths of Si$_A$-Si$_B$ are ~2.33-2.35 Å, and of Si$_B$-H are ~1.52 Å. The buckling distance of both structures is found to be 0.71 Å. These values agree with previous findings for hydrogenated silicene [3, 40, 42, 48, 52, 53, 56]. In order to find the favored coupling of the magnetic moments, three different magnetic types are considered: antiferromagnetic (AFM), FM, and NM states. The spin polarization energy ($\Delta E_{SP}$ = $E_{NM}$ − $E_{FM}$) and the FM energy ($\Delta E_{FM}$ = $E_{AFM}$ − $E_{FM}$) are calculated. It is found that $\Delta E_{SP}$ and $\Delta E_{FM}$ are 5.86 (8.38) eV and 0.264 (0.427) eV for hH-4-ZT-SiNF (hH-3-ZH-SiNF), respectively, indicating that half hydrogenation of SiNFs leads to a FM state. Also, the net magnetic moment of hH-4-ZT-SiNF is 18.0 μ$_B$, and hH-3-ZH-SiNF is 27.0 μ$_B$, which are considerably larger than that of bare ZT-SiNF (3.0 μ$_B$) and 3-ZH-SiNF (0.0 μ$_B$). To visualize spin distributions on half-hydrogenated SiNFs, the spin densities of hH-4-ZT-SiNF and hH-3-ZH-SiNF are plotted in Fig. 5(b). As can be seen, the systems provide FM spin ordering due to the parallel alignment of spin moments. The magnetic moments for Si$_A$, Si$_B$, and H which are marked in Fig. 5(b) are calculated to be 0.923, −0.011, and 0.088 μ$_B$, respectively. It is clear that the induced magnetic moments are mainly localized on unsaturated Si$_A$ atoms (92%), while the H atoms above Si$_B$ atoms carry small magnetic moments (8%), showing that the unpaired *3p$_z$* electrons in the unsaturated Si$_A$ atoms have the highest contribution to the magnetism. Since the *3p$_z$* electrons on Si$_A$ are delocalized to some extent, a long-range FM exchange coupling between the $p_z$ electrons of different Si$_A$ atoms exists due to the extended *p-p* interactions, giving rise to a significant collective character.

If the hydrogenation of SiNFs continues, the unsaturated Si$_A$ atoms will be exposed by H atoms. Therefore, the *3p$_z$* orbitals of the unsaturated Si$_A$ atoms will be saturated by *1s* orbitals of H atoms, causing the spin states to be suppressed. Similar to silicane

[3], fully hydrogenating (fH) of SiNFs where all $Si_A$ and $Si_B$ atoms are saturated by H atoms, quenches the magnetism. It is found that the fH-4-ZT-SiNF and fH-3-ZH-SiNF become NM wide band gap semiconductors with energy gaps of 3.16 and 2.8 eV, respectively.

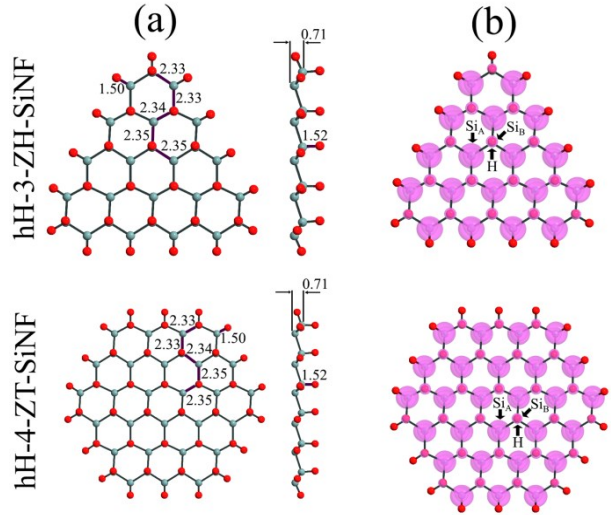

Fig. 5 (a) Optimized geometric structures (b) the corresponding spin-charge density distribution of hH-3-ZH-SiNF and hH-4-ZT-SiNF.

### 3.4. The total spins moment of bare and half-hydrogenated SiNFs

Next, we develop a general rule for the total spin moment of nanoflakes as a function of nanoflakes size. It is found that the total magnetic moment of bare SiNF satisfies $M_{bare} = N_A - N_B$, where $N_A$ and $N_B$ are the numbers of $Si_A$ and $Si_B$ atoms in the nanoflakes. Among the four studied SiNF structures, only ZT-SiNF has a different number of Si atoms in each sublattice ($N_A \neq N_B$), therefore, the magnetic moment of 4-ZT-SiNF, as an example, with $N_A = 18$ and $N_B = 15$ is equal to $M_{bare} = 3.0$ $\mu_B$. However, the total magnetic moments for the rest of the structures are calculated to be $M_{bare} = 0.0$ $\mu_B$. For instance, the magnetic moment of 3-HT-SiNF with $N_A = 27$ and $N_B = 27$ is equal to 0.0 $\mu_B$. As found before, a single H can boost the SiNF's spin moment by 1.0 $\mu_B$. Consequently, the induced spin moments of half-hydrogenated SiNF where all $Si_B$ atoms are saturated and $Si_A$ atoms keep unsaturated is $M_{hH} = M_{bare} + N_B \times 1$. As calculated before, the total spin moment for hH-4-ZT-SiNF is 18.0 $\mu_B$, and for hH-3-ZH-SiNF is 27.0 $\mu_B$. More careful examination of these values reveals that the total magnetic moment for half-hydrogenated zigzag triangular and hexagonal triangular SiNFs can be expressed as $M_{hH} = (5N + N^2)/2$ and $M_{hH} = 3N^2$, respectively, where $N$ is the nanoflakes size. Fig. 6(a) represents the variations of total spin moment of half-hydrogenated zigzag triangular and hexagonal SiNFs as a function of nanoflakes size. The total spin moments of hydrogenated SiNFs are notably larger than that of bare SiNFs and almost increase squarely with the nanoflakes size. Compared to graphene, it is found that the total spin moment of hH-SiNFs is two times larger than that of hH-GNFs [69, 70].

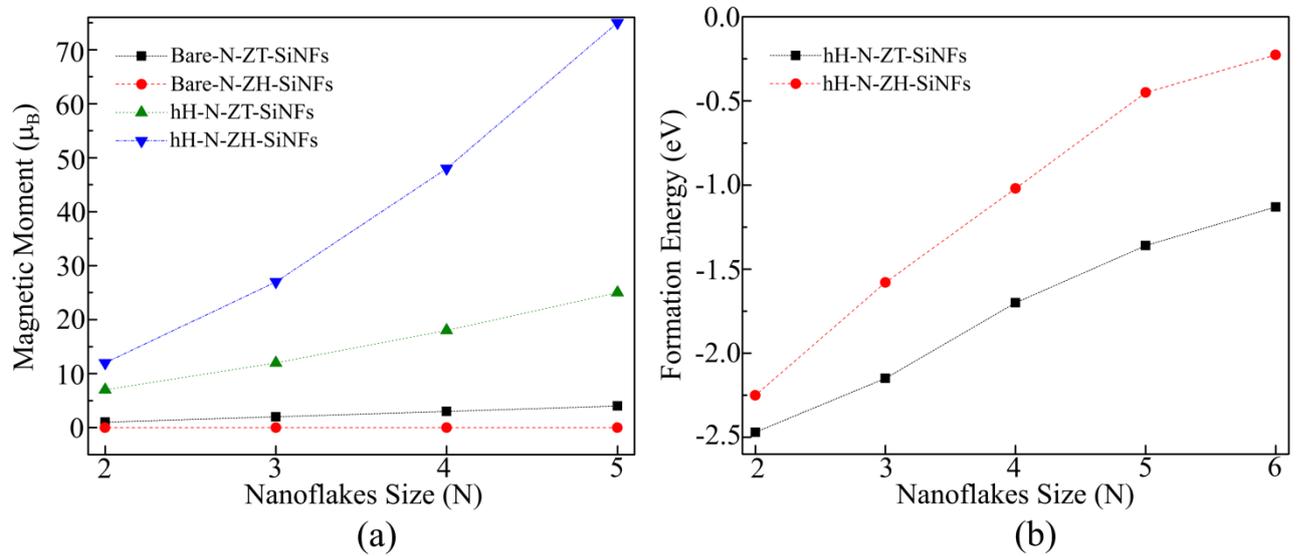

Fig. 6 (a) The magnetic moment changes *versus* nanoflakes' size for Bare-N-ZT-SiNFs, Bare-N-ZH-SiNFs, hH-N-ZT-SiNFs, and hH-N-ZH-SiNFs (b) the formation energy variations of hH-N-ZT-SiNFs and hH-N-ZH with nanoflakes' size.

### 3.5. The stability of half-hydrogenated SiNFs

The stability of hydrogenated nanoflakes is of great significance for practical applications. To address this important factor, the structural stabilities of the hH-N-ZT-SiNF and hH-N-ZH-SiNF configurations with different nanoflakes size ($N$) are evaluated using formation energy ($E_F$) which is

$$E_F = [E_{hH-SiNF} - (E_{SiNF} + n_H \times E_{H_2}/2)]/n_H$$

Here, $E_{hH-SiNF}$ and $E_{SiNF}$ are the total energies of half-hydrogenated and bare SiNF, respectively. $E_{H2}$ denotes total energy of an isolated $H_2$ molecule, and $n_H$ represents the difference of the number of hydrogen atoms between two systems and is equal to $Si_B$ atoms. The variations of formation energies of hH-N-ZT-SiNF and hH-N-ZH-SiNF configurations as a function of nanoflakes size are plotted in Fig. 6(b). The calculated formation energies for hH-4-ZT-SiNF and hH-3-ZH-SiNF are found to be –1.7 and –1.58 eV, respectively. Based on the definition, the more negative value of $E_F$ is, the more stable the nanoflake becomes. As can be seen, all the formation energies are negative, showing that the half-hydrogenated SiNFs are stable. In addition, the formation energies increase monotonically with increasing nanoflakes size, indicating that small size nanoflakes are experimentally more favorable to form.

### 3.6. DOS and PDOS of bare and hydrogenated SiNFs

In order to gain more insight into the origin of magnetic behavior of hydrogenated SiNFs, the projected density of states (PDOS) of bare, single-hydrogenated, and half-hydrogenated ZT-SiNF and ZH-SiNF are calculated, as shown in Fig. 7.

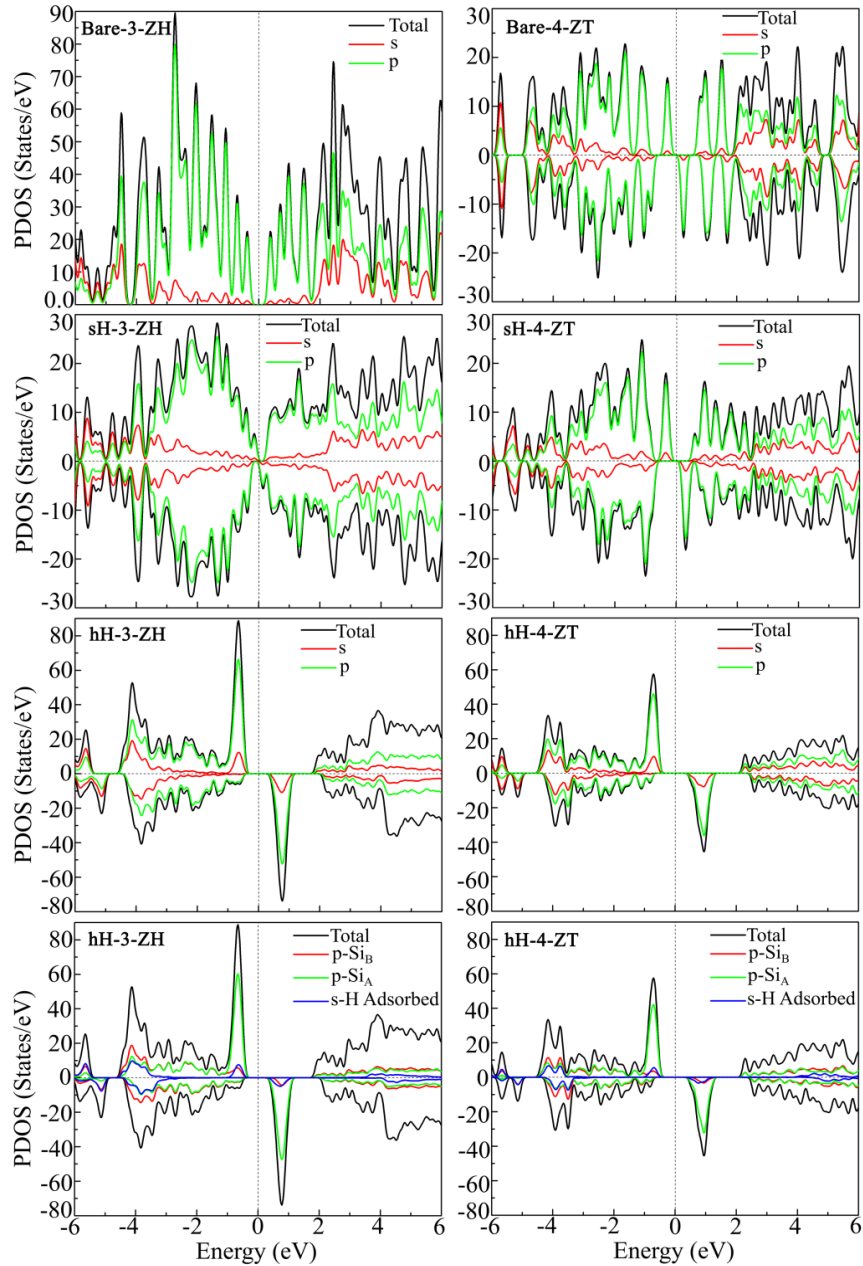

Fig. 7 Spin-polarized PDOS of Bare-3-ZH-SiNF, sH-3-ZH-SiNF, hH-3-ZH-SiNF, Bare-4-ZT-SiNF, sH-4-ZT-SiNF, and hH-4-ZT-SiNF, with the spin-up (spin-down) component on the positive (negative) axis.

In agreement with aforementioned findings, the 3-ZH-SiNF is a NM semiconductor with a band gap of 0.76 eV. However, the 4-ZT-SiNF behaves as a magnetic semiconductor with direct band gaps of 1.05 eV in the spin-up, and 1.07 eV in the spin-down states. Although the silicon atom (Ne $3s^2$ $3p^2$) has four valence electrons in the flake, the contributions of s orbitals are negligible near the Fermi level. It is clear that the magnetic behavior of triangular SiNF originates from the $p$ electrons located near the Fermi level. The value of the prominent peak close to the Fermi level at $E \sim$

−0.26 (0.26) eV for spin-up (-down) channel is 15.9 (−16.9) states/eV. The spin-up and spin-down states show the similar distribution in reciprocal space. However, the located energy of the former is slightly lower than that of later in high energy levels. Upon adsorption of a single H atom on $Si_B$ of 4-ZT-SiNF, the *p* orbital contributions of Si atoms close to the Fermi level in both spin-up and spin down channels are strengthened, confirming a stronger magnetism in the nanoflakes. Interestingly, the adsorption of a single H atom onto $Si_B$ of 3-ZH-SiNF introduces new spin-up and spin-down states near the Fermi level, mainly attributed to the *3p* electrons of $Si_A$ atom. Accordingly, the total magnetic moments of both cases are improved by 1.0 $\mu_B$. As stated before, the non-magnetic and ferrimagnetic behavior of bare hexagonal and triangular SiNF are transformed to very strong FM behavior after half-hydrogenation of the SiNFs. For hH-4-ZT-SiNF, the value of the peak close to the Fermi level at $E \sim -0.71$ (0.94) eV for spin-up (-down) channel is increased to 57.5 (− 45.6) states/eV which is much greater than of bare 4-ZT-SiNF. The contributions of *p* and *s* orbitals in these peaks are 79% and 17%, respectively. A closer look at the PDOS of hH-4-ZT-SiNF indicates that the *p* orbitals of $Si_A$ atoms have the highest (72%), and the *s* orbitals of adsorbed H atoms (10%), and the *p* orbitals of $Si_B$ (7%) have second and third contributions to the magnetism, respectively. It is important to mention that hH-4-ZT-SiNF is a FM semiconductor with direct band gaps of 2.87 eV in the spin-up, and 1.37 eV in the spin-down states. A strong peak of 88.1 (− 73.9) states/eV is induced at E ~ − 0.66 (0.76) eV for spin-up (-down) channel through half-hydrogenation of 3-ZH-SiNF, which is much greater than of hH-4-ZT-SiNF, indicating that the half-hydrogenated hexagonal SiNF can introduce a very large spin. These peaks mainly contribute to the *p* (74%) and *s* (14%) orbitals.

### 3.7. Spin switch devices based on hydrogenated SiNFs

In order to explore the potential applications of hydrogenated SiNFs in spintronics, simple spin switch devices have been proposed. In these structures, half-hydrogenated and fully hydrogenated 4-ZT-SiNFs are sandwiched between two conducting bare 5-ZSiNRs. Fig. 8 shows the geometric structures of spin switch devices and the spin density distributions of sandwiched nanoflakes. If all $Si_B$ atoms of the nanoflake are saturated with H atoms, the spin-up states with similar alignment appear on the half-hydrogenated SiNFs, as shown in Fig. 8(a). It suggests that this model is an ideal transport channel only for the spin-up electrons even when considering antiparallel spin orientation between two edges of ZSiNFs. Despite this, hydrogenation of all $Si_A$ atoms of SiNF makes it a perfect transport channel for only spin-down electrons, as seen in Fig. 8(b). In this case, all the spin-down states with same spin configuration appear on the half-hydrogenated SiNFs, which block the transmission of spin-up electrons coming from conducting ZSiNR. As stated earlier, the full hydrogenation of the nanoflakes quenches their magnetism. Therefore, a spin switch device made of the fully hydrogenated SiNFs block the transmission of both spin-up and spin-down electrons, as indicated in Fig. 8(c). In contrast to a bare ZSiNR with hydrogenation that transports both spin-up and spin-down electrons along the edges [71]. The hydrogenation coverage can be manipulated due to its reversibility [36] and controllability [37, 38]; therefore, it is possible to tune the transport and magnetic properties, opening new gates into novel spintronic circuit devices.

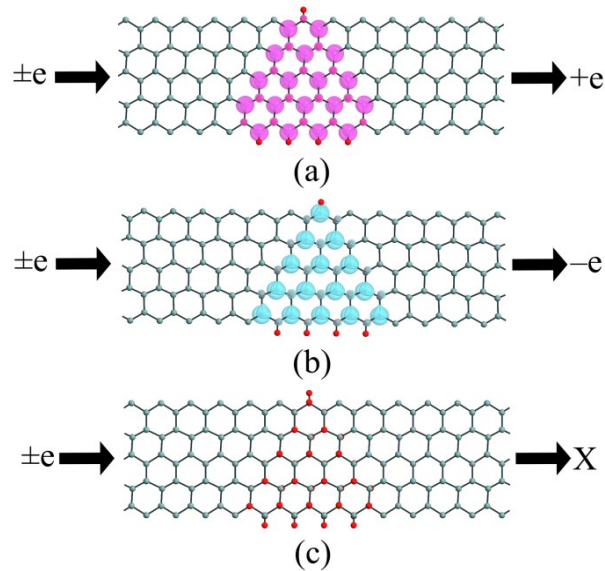

Fig. 8 Geometric structures of (a) the hH-4-ZT-SiNF with all $Si_B$ atoms covered by H atoms, (b) the hH-4-ZT-SiNF with all $Si_A$ atoms covered by H atoms, and (c) the fH-4-ZT-SiNF with all $Si_A$ and $Si_B$ atoms covered by H atoms, sandwiched between two bare 5-ZSiNRs. The spin density distributions of ZT-SiNFs are only plotted. The pink and blue regions correspond to the isosurfaces of spin-up and spin-down states, with the isosurfaces value of 0.012 a.u.

**Conclusions**

In summary, first-principles calculations based on DFT method are employed to study the structural, electronic, and magnetic properties of SiNFs with different configurations. It is found that while both zigzag-edged hexagonal and triangular SiNFs show semiconducting behavior, the former is NM and the latter has ferrimagnetic character. Hydrogenation is accepted as an effective approach to engineer the electronic and magnetic properties of SiNFs owing to its controllability and reversibility. Upon adsorption of a single H atom on a Si atom of a SiNF, the quasi $\pi$ bond between saturated and unsaturated Si atoms is broken because of a strong $\sigma$ bond forms between saturated Si and H atoms. It causes the $p_z$ electrons of the unsaturated Si atom to be unpaired and delocalized. As a consequence, the magnetism in SiNFs is augmented by 1.0 $\mu_B$. Increasing the hydrogenation coverage on SiNF to half of the Si atoms gives rise to a very large spin moment that is squarely proportional to the size of the nanoflakes. It is discovered that half-hydrogenated SiNFs present a long-range FM order because of the extended *p-p* interactions of adjacent Si atoms. Moreover, full hydrogenation of SiNF quenches the magnetism due to the saturation of all the quasi $\pi$ bonds. Finally, simple switch devices are proposed to show the potential application of silicene to tune the transport properties by harnessing the hydrogenation coverage on SiNFs.

## Acknowledgements

This work was supported in part by the Florida Education Fund's McKnight Junior Faculty Fellowship.